\documentstyle[preprint,prl,aps,epsf]{revtex} 
\begin{document}
\bibliographystyle{plain}
\title{An Analytical Approach to the Protein Designability Problem.}
\author{Edo L. Kussell, Eugene I. Shakhnovich }
\address{Harvard
University, Department of Chemistry and Chemical Biology \\
12 Oxford Street, Cambridge, MA  02138 }
\date{\today}
\maketitle
\begin{abstract}
We present an analytical method for determining the designability of protein
structures. We apply our method to the case 
of two-dimensional lattice
structures, and give a systematic solution 
for the spectrum of any structure. Using
this spectrum, the designability of a 
structure can be estimated. We outline a
heirarchy of structures, from most to 
least designable, and show that this
heirarchy depends on the potential 
that is used. \\

\end{abstract}

\def \N {\cal{N}}
\def \si {\sigma _i}
\def \csi { \{ \si \} }
\def \eio {e^{i \omega C} d \omega }
\def \lp {\lambda_+}

\def \lm {\lambda_-}
\def \ep {e^{i \omega}}
\def \em {e^{-i \omega}}
\def \etp {e^{2 i \omega}}
\def \etm {e^{-2 i \omega}}
\def  \natstr {\{r_i^0\}}
\def  \any_str {\{r_i\}}
\def   \seq    {\{ \sigma _i \}}

The protein design problem asks which and how many 
amino-acid sequences fold into a given protein structure. 
This problem, having obvious practical
and evolutionary significance,
attracted considerable attention and effort
of experimentalists \cite{Mayo} 
and theorists
\cite{PNAS,Pande:94,DES_REV,NEC}. 
In particular, Tang and coworkers carried out a thorough study
of the design problem for a simple square and cubic lattice 
model of proteins \cite{NEC}. These authors enumerated
all sequences made of two types of monomers and all conformations
for 2-dimensional 36-mers and 
3-dimensional 27-mers. This calculation
provided
a density of states 
in sequence space for the studied models, i.e.
 how many sequences
have a given energy in a given conformation. 
For each conformation, 
the ``designability,'' which is defined
as the number
of sequences that have this conformation as a ground state,
was determined in \cite{NEC}. Interestingly, 
it was found that designability
varied from conformation to conformation. Furthermore, 
the structures
that possess certain degrees of symmetry turned out to be
most designable \cite{NEC}. This result 
tempts one to speculate that
the factor of designability 
may be the cause of symmetries observed in real proteins. \\
\indent The physical reason for the observed
relationship between geometric properties of 
conformations and their designability remained unclear.
In particular, it is important to evaluate how robust
is this relationship with respect to choice of 
lattice model, model protein ``alphabet,'' and
interaction parameters between 
model aminoacids. Recent
numeric studies \cite{Goldstein:99} 
and \cite{Ejtehadi:98}
showed that the set of the most designable 
structures does indeed depend
on the 
potential and aminoacid alphabet used. \\
\indent A deeper understanding  of the designability problem
requires an analytical solution that may shed light on 
the general geometric features that make structures designable.
Such a solution will facilitate an
 informed connection between
model results and real proteins. \\
\indent In this Letter  we present a statistical mechanical analysis
of designability, and give an 
analytical solution to the problem on a two-dimensional 
lattice. 
Our analysis is based on the 
analogy between protein design and a class of statistical-mechanical
spin models established in earlier works 
\cite{PNAS,DES_REV}. \\
\indent 
We use the standard lattice 
 model to represent protein conformations. 
 On a lattice, 
a structure is defined as a self-avoiding walk of 
length $N$. 
The energy of a sequence of residues $\{ \sigma _i \}$ 
in a  structure with coordinates of the monomers
$\{r_i \}$ is 
\begin{equation}
H( \{ \sigma_i \}, \{ r_i \} ) = \sum_{i \, < \, j}^{N} 
U(\sigma _i , \sigma _j ) 
\Delta (r_i , r_j)
\label{eq:HAM}
\end{equation}
$\Delta (r_i , r_j)$ is defined to be 1 
when $r_i$ and $r_j$ are nearest-neighbors on the 
lattice and not sequence neighbors along the chain. 
For real proteins the definition of a contact requires a 
choice of a cut-off distance.
The folding potential  
is given by a symmetric matrix $U(\alpha, \beta)$ 
whose entries are the pairwise interaction energy 
between aminoacid residues of types $\alpha$ and $\beta$. \\
\indent In any instance of the design problem, 
the structure $\natstr$
is given  (we will refer to
this structure as the ``target conformation'' for design), 
and we are asked to find all sequences whose 
native (folded) state is $\natstr$. 
The number $\N$ of such sequences 
is the {\it designability} \rm of the structure. 
Since the residues of a protein are tightly 
packed in space, only maximally compact lattice 
structures are generally studied. On a square 
lattice these correspond to self-avoiding walks 
that completely fill a square. The folding 
potential $H$ is therefore assumed to have a 
non-specific attractive term (which we have omitted), 
that favors compact conformations. \\
\indent In order for a sequence to fold and 
to be stable in the desired target
conformation it must have lowest energy 
in this conformation compared to its energy in
all alternative conformations. 
The spectrum of energies of alternative conformations
is qualitatively similar to that of a random heteropolymer,
i.e. it consists of a continuum part, with a high density of states whose
lowest energy is $E_c$, and of a 
few discrete energy levels lying below $E_c$ \cite{NATUR,B.W.}. 
Importantly, the value of $E_c$ is self-averaging \cite{NATUR}, i.e it
depends on amionacid composition rather than on a particular
sequence \cite{PRLF,DES_REV}. While this description is strictly applicable
to three-dimesional heteropolymers (see though \cite{Grosberg:96}, it was
shown in \cite{dinner:94} that it qualitatively applies to two-dimenisonal
case.\\
\indent Thus the issue of 
designability is reduced to the question of how many
sequences with a given aminoacid composition 
can fold into a given conformation
with an energy $E \ll E_c$.  Each structure has a sequence-space 
energy spectrum given by the energy density
\begin{equation}
n(E, \Delta) = \sum _{\{ \si \}} \delta (H(\{ \si \}, \Delta) - E)
\label{eq:spectrum}
\end{equation}
The above expression corresponds to a system of $N$ spins $\{ \sigma _i \}$ 
confined to an interaction geometry dictated 
by $\Delta ( r_i , r_j )$, with $\sigma _i \in 1 \ldots q$, 
where $q$ is the number of possible 
residue types. 
In this language the 
aminoacid composition of a sequence is equivalent to total
magnetization 
and is denoted by the letter $C$, 
which is in general a $q$-dimensional vector. 
The probability $p(E, C)$ that a 
sequence $\{ \si \}$ of composition $C$, will fold 
into a structure $\Delta$, where $E = H(\{ \si \}, \Delta )$, is
\cite{Gutin}
\begin{equation} 
p(E, C) = 
 \exp \left( \frac{-1}{\sqrt{4 \pi N \log \gamma}} \exp \left( \frac{E - E_c}{T_c} \right) \right)
\end{equation}
where $T_c$ is the temperature of the
thermodynamic ``freezing'' transition in 
random heteropolymers having composition $C$,
 and $\gamma$ is the effective number 
of conformations per monomer \cite{Gutin}. These 
properties of $E_c$ render it a valuable link 
between sequence and structure space. If we know 
the spectrum $n(E, \Delta, C)$ of a structure 
$\Delta$ over all sequences of fixed composition C, 
we can compute its expected designability as follows:
\begin{equation}
\cal{N}(\Delta) = \sum_{E, C} p(E, C) 
n(E, \Delta, C) \label{expectDesign}
\end{equation}
\indent On a two-dimensional square lattice, maximally 
compact structures have a property that 
makes the spectrum calculation $n(E, \Delta, C)$ 
analytically 
tractable. If the ends of the chain are prescribed 
to be neutral - they do not interact with any 
monomers - then the entire structure can be decomposed 
into a direct product of one-dimensional loops 
and strands. A loop is a collection of $S$ monomers 
in which monomer $i$ is in contact with monomer 
$i+1$, for $1 \leq i \leq S - 1$, and monomer 1 
is in contact with monomer $S$; if monomer 1 and 
monomer $S$ are not in contact, we call the 
collection a strand. Figure 1 shows the decomposition 
of a particular 36-mer into strands and loops (ends have 
been allowed to interact in this figure).
\marginpar{Fig.1}
This decomposition derives from the restriction 
that nearest neighbors 
along the chain are not in contact, leaving each 
monomer in the interior of a structure exactly 2 
contacts to make. Since this excludes branched 
systems of contacts, the only possibilities left 
are loops and strands. Monomers lying on the square 
border of a structure can make only one contact 
because the chain must occupy two of the three 
neighboring sites. Strands, therefore, must begin 
and end on the border of a structure, and loops 
must lie entirely within it. Since the corners 
of a structure cannot interact, a square structure 
of length $N$ has $4 \sqrt{N} - 8$ interacting 
border sites. Thus, there must be $2 \sqrt{N} - 4$ 
strands in each structure. Including the ends of 
the chain would add at most 2 branched systems, or 
would change the number of strands by at most 2. 
While the number of strands is largely the same for all 
structures, the distribution of lengths 
of strands is structure-dependent, and
will prove to be the 
determinant of designability. \\
\indent We restrict this discussion to 
systems with two types of aminoacids, which we denote
+1 and -1, 
although the general methodology can certainly 
be applied to larger alphabets. Following the analogy
between protein design and spin models \cite{PNAS,DES_REV}, 
aminoacid types will be called ``spins'' in what follows.
The interaction energies are taken to be 
$U(1, 1) = a$, $U(-1, -1) = b$, 
and $U(1, -1) = U(-1, 1) = d$. The 
general aim is to compute $n(E, \Delta, C)$, 
which can be deduced from the partition 
function at composition C, over all 
sequences $\csi$:
\begin{equation}
Z(C) = \sum_{\csi} e^{- \beta H \csi } \delta ( C - \sum_i \si )
\end{equation}
Note that composition in the 2-spin 
case is a scalar corresponding to the 
net magnetization. The delta function 
can be expressed as a Fourier integral, 
and if we define the parition function 
for the $j$-th loop or strand to be
\begin{equation}
Z_j (\omega) = \sum_{\csi _j} e^{- \beta H \csi - i \omega \sum \si}
\end{equation}
where $\csi _j$ is the set of all 
spins in the $j$-th loop or 
strand, we can write
\begin{equation}
Z(C) = \int \eio \prod_{j = 1}^{s + l} Z_j (\omega) \label{Z(C)}
\end{equation}
where $s$ and $l$ are the number of strands 
and loops, respectively. $Z_j (\omega)$ can 
be computed by the transfer matrix method; 
we define the transfer matrix $T$ as
\begin{equation}
T = \left( \begin{array}{ll}
        A e^{-i \omega} & D \\
        D               & B e^{i \omega} \
\end{array} \right)
\end{equation}
where $A = e^{-\beta a}$, $B = e^{-\beta b}$, 
and $D = e^{-\beta d}$. If $\lp$ and $\lm$ 
are the eigenvalues of $T$, we have 
for a loop, $Z_j (\omega) = \lp ^{n_j} + \lm ^{n_j}$, while for a strand
\begin{equation}
Z_j (\omega) = \sum _{\sigma _1 , \: \sigma _{n_j}} 
e^{-(\sigma _1 + \sigma_{n_j}) i \omega / 2} \langle \sigma _1 | T^{n_j - 1} | 
\sigma _{n_j} \rangle
\end{equation}
where $n_j$ is the length of the $j$-th loop or strand. 
The expression for strands is more complicated than 
that for loops because the first and last monomers 
in a strand have to be summed over separately. 
By diagonalizing $T$, the following expression 
for strands can be derived:
\begin{eqnarray}
Z_j (\omega) = \left[ (2 D - A - B) (\lp ^{n_j - 1} -
\lm ^{n_j - 1} ) +  \nonumber \right. \\ \left. (e^{i \omega} + e^{-i \omega})
(\lp ^{n_j} - \lm ^{n_j}) \right] / (\lp - \lm) \label{Zstrand}
\end{eqnarray}
The eigenvalues can be calculated, and one observes 
that if $D^2 = A B$, we get $\lp = A \em + B \ep$, while $\lm = 0$. 
This situation corresponds to any 
potential for which $d = (a + b)/2$. We will 
call this the symmetric potential, because any 
such potential is the same as $a = 1$, $b = -1$, $d = 0$. Under the 
symmetric potential, $Z_j (\omega) = \lp ^{n_j}$ for 
loops, and $Z_j (\omega)$ for strands simplifies 
considerably to
\begin{equation}
Z_j (\omega) = \lp ^{n_j - 2} (\sqrt{A} \em + \sqrt{B} \ep )^2 
\label{Zstrandsimp}
\end{equation}
If $N$ is the chain length of a structure, 
the total partition function (\ref{Z(C)}) 
for the symmetric potential now becomes
\begin{equation}
Z(C) = \int \eio \lp ^{N - 2 s} (\sqrt{A} \em + \sqrt{B} \ep )^{2 s}
\end{equation}
Using the binomial expansion, the integral can be calculated exactly 
and the density of states, $n(E, C)$ obtained. The following is a plot of $n(E,
C)$ using 
$a = -1$, $b = 1$, N = 36, and C = 0:
\marginpar{Fig.2}

\indent Under the symmetric potential, a 
structure's spectrum depends only on its length, 
composition, and the number of strands. Since 
these three are the same for each structure, 
the spectra of all structures are identical 
(up to end-effects). This result implies that 
all structures have the same expected designability, 
with fluctuations only due to the effects of 
interacting ends and to any sequence-dependent 
terms of $E_c$. For large $N$ these terms 
will be insignificant. \\
\indent The symmetric potential is a singular 
case in that structural features play no role in 
designability. It has previously been identified 
in \cite{Ejtehadi:98} as 
an important special case for design. 
We now show how designability 
becomes sensitive to structure under 
perturbations of this potential. \\
\indent Under the symmetric potential, one eigenvalue 
is zero. If we add a small perturbation $u$ to 
the potential, so that $a' = a + u$, $b' = b$, 
and $c' = (a + b)/2$, we would expect one 
eigenvalue still to be small compared to 
the other. Under a perturbed potential, 
then, $\lm / \lp \ll 1$. Starting from the 
general expression for strands (\ref{Zstrand}), 
we observe that if $\lm < \lp$, we have two 
geometrical series in $R = \lm / \lp$, and we 
can rewrite it as:
\begin{equation}
Z_{j}(\omega) = \lp ^{n_j - 2} \Theta 
\left[ 1 + \sum_{k=1}^{n_j - 2} R^k - 
(\ep + \em )\frac{\lm}{\Theta} R^{k - 1} \right]
\end{equation}
where $\Theta = A \etm + B \etp + 2 D$. 
The corresponding expression for loops is
\begin{equation}
Z_{j}(\omega) = \lp ^{n_j} ( 1 + R^{n_j})
\end{equation}
Using these expressions in equation (\ref{Z(C)}) 
for $Z(C)$ results in a product of the loop and 
strand $Z_j$'s. We can expand this product in $R$, 
and obtain the following kinds of terms: 
The principal contribution  - $\lp ^{N - 2 s} \Theta ^s$ - 
(zero-order in R) is a
structure-independent term corresponding to the symmetric
potential. Terms coming from a 
single loop of length $n_j$ look 
like $\lp ^{N - 2 s} \Theta ^{s} R^{n_j}$. 
Terms coming from a single strand of length $n_j$ are:
\begin{equation}
\lp ^{N - 2 s} \Theta ^{s} \left[R^k - (\ep + \em )\frac{\lm}{\Theta} R^{k - 1}
\right] \label{strandpert}
\end{equation}
where $1 \leq k \leq n_j - 2$. Strands of 
length 2 contribute no terms. Thus, strands have terms 
at all powers of $R$ less than $n_j - 1$, while 
loops only perturb at $R^{n_j}$. Since the smallest 
possible loop is of length 4, strands of 
lengths greater than 2 are much more important 
to the structure of the spectrum than 
loops of any length. \\
\indent To calculate a perturbed spectrum, 
one still has to back-transform the Fourier 
variable $\omega$ in each term. Setting 
$U = e^{-\beta u}$, the perturbed eigenvalues 
must satisfy $\lp + \lm = A U \em + B \ep$ 
and $\lp \lm = A B (U - 1)$. The principal contribution 
to the partition function is
\begin{equation}
Z_{principal}   = \int \lp ^{N - 2 s} \Theta ^s \eio
\end{equation}
which can be computed by binomial 
expansions of $\lp ^{N - 2 s}$. The only 
difficulty in doing so comes from 
odd powers of a square-root term in the expression
for $\lp$ which would normally lead to a power 
series expansion rather than a finite 
binomial expansion. The approximation 
we make in our computation is to 
replace the odd powers of the 
square root with the next even 
number, so that we have terminating 
binomial expansions at all powers. 
To compute a perturbation of the 
form $\lp ^{N - 2 s} \Theta ^s R^k$ 
we use $R = A B (U - 1) / {\lp ^2}$ 
and obtain: 
\begin{equation}
Z_{perturb}  = \int \lp ^{N - 2 s - 2 k} \Theta ^{s} A^k B^k (U
- 1)^k \eio
\end{equation}
which is similar to $Z_{principal}$ except 
for another binomial that can be expanded. Computing 
perturbations of any order is therefore relatively 
straightforward. \\
\indent We performed computations 
to check our theory 
with a full enumeration of sequences and structures 
that was published by Tang and coauthors \cite{NEC}. 
These authors used a 
potential ($a = -2.3$, $b = 0$, $d = -1$) which can 
be rescaled to the perturbed potential 
$a = -1$, $b = 1$, $d = 0$, $u = -0.3$. 
The following is a log plot of the 
principal (structure-independent) 
contribution to the spectrum of a 36-mer:
\marginpar{Fig.3}

This lowest-order approximation of the 
density of states bears a striking 
resemblance to the spectrum calculated 
by  full enumeration for 3D structures (C.Tang private communication).
Unfortunately our request to provide sample densities
of states for 2D structures had not been granted by
the authors of \cite{NEC}. 
Qualitative features such as the shape and the 
anomalous speckeled pattern emerge from our theory. 
The pattern of the spectrum is essentially a splitting of the 
states of the symmetric potential due to the 
perturbation $u$. \\
\indent The expected designability (\ref{expectDesign}) 
depends on the value of $E_c$. The 
following is a plot of the change in 
expected designability due to the first-order 
strand perturbation as a function of $E_c$. Note that 
the logarithm of negative numbers 
was taken in absolute value and plotted 
as a negative value. 
\marginpar{Fig.4}
\indent The change 
in the expected designability of the 
structure due to this perturbation is negative 
for a large range of $E_c$ in the lower part
of the spectrum. Since strands of lengths greater 
than 2 will contribute a first-order 
perturbation, the more such strands in a 
structure, the less designable it will be. 
The most designable structures will, therefore, 
contain the maximum number of strands of 
length 2. This is the major effect determining 
designability; adding higher-order 
terms does not change this result. The effect 
which 2-strands have on designability is 
due to the condition $u < 0$. One can also 
perturb with $u > 0$ and obtain the opposite 
result: 2-strands are not desirable. All the plots shown here 
correspond to composition $C = 0$. 
Similar plots are obtained over a large 
range of $C$, with significant 
qualitative differences for $| C |$ 
close to $N$, i.e. as the sequence 
approaches a homopolymer. Our 
conclusion remains unchanged by 
these differences because at 
extreme compositions we expect 
very few sequences to design 
any given structure. \\
\indent Deciding whether or not a strand 
of length $n$ is better for designability 
than a strand of length $n + 1$  requires looking
at the change in expected designability due to 
higher order perturbations. We find that up to strands 
of lengths 5, the shorter the strand, the more designable 
the structure. For longer lengths, differences in designability 
become very small, and resolution of these differences 
requires precise knowledge of $E_c$. \\
\indent We have presented a general, 
analytical approach to the protein 
design problem, and by applying it 
to the case of two-dimensional structures, 
we have shown how it effectively recognizes 
features that make some structures 
much more designable than others. 
This allowed us to 
identify a heirarchy of 
designability based on strand lengths. 
Future work should generalize 
these results to larger alphabets (which leads 
to a $q$-th degree polynomial in $\lambda$), and to 
lattices of dimension three (which leads to highly 
branched systems of contacts).\\
\indent We thank A.Grosberg for useful comments. The work was supported
by NSF scholarship (to EK) and NIH grant GM52126 (to ES).

\pagebreak
\section*{Figure Captions}

\noindent FIG. 1. Strand/Loop decomposition for a 36-mer \\ \\

\noindent FIG. 2. Symmetric potential spectrum of a 36-mer \\ \\

\noindent FIG. 3. Principal contribution to 36-mer spectrum with u = -0.3 \\ \\

\noindent FIG. 4. Change in expected designability due to 1st-order perturbation

\pagebreak

\pagestyle{empty}

\begin{center}
\par \leavevmode\def\epsfsize#1#2{2.0 #1}\epsffile{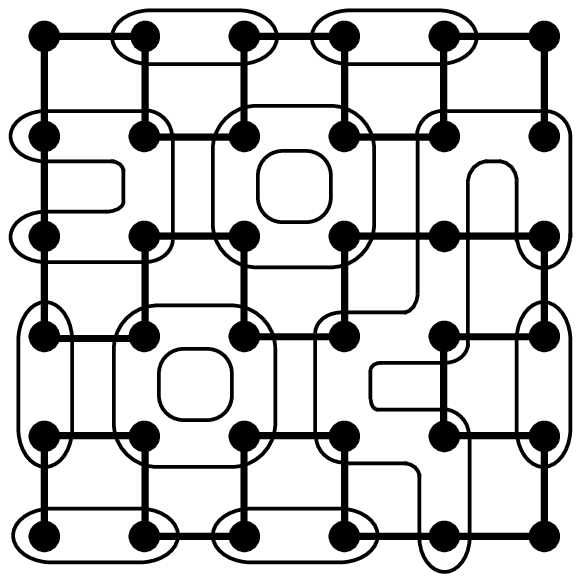}
\end{center}

\vspace{10pt}
\begin{flushright}
Kassel and Shakhnovich Fig.1 
\end{flushright}

\pagebreak

\begin{center}
\par \leavevmode\def\epsfsize#1#2{2.0 #1}\epsffile{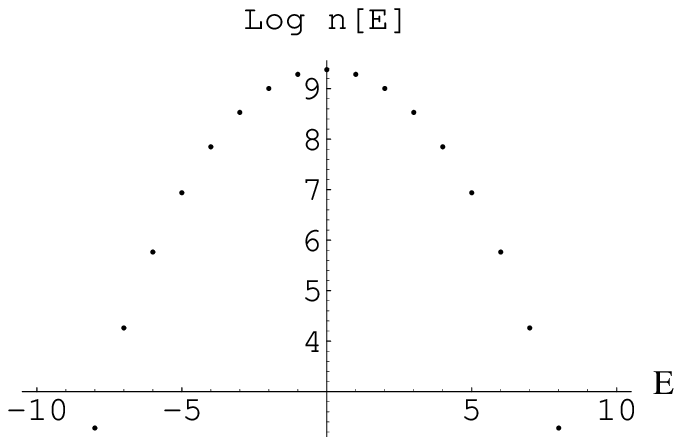}
\end{center}

\vspace{10pt}
\begin{flushright}
Kassel and Shakhnovich Fig.2 
\end{flushright}

\pagebreak

\begin{center}
\par \leavevmode\def\epsfsize#1#2{2.0 #1}\epsffile{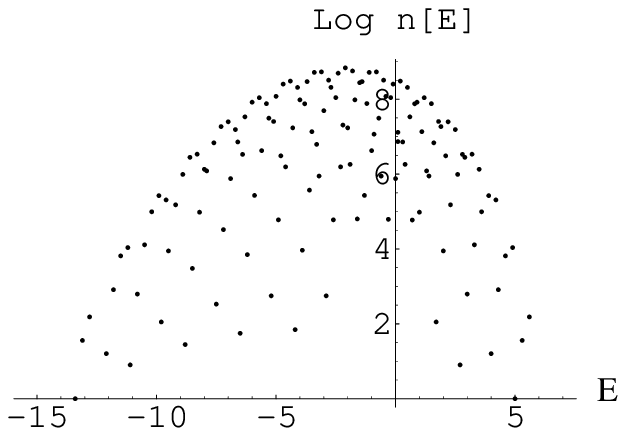}
\end{center}

\vspace{10pt}
\begin{flushright}
Kassel and Shakhnovich Fig.3 
\end{flushright}

\pagebreak

\begin{center}
\par \leavevmode\def\epsfsize#1#2{2.0 #1}\epsffile{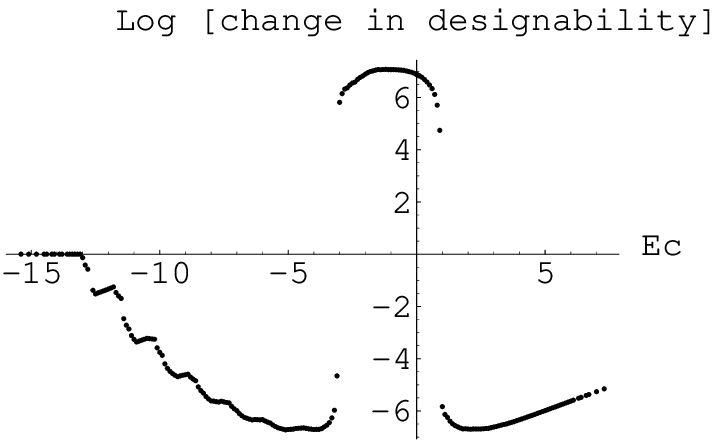}
\end{center}

\vspace{10pt}
\begin{flushright}
Kassel and Shakhnovich Fig.4 
\end{flushright}

\end{document}